\documentclass[aps,prb,reprint,amsmath,amssymb,floatfix]{revtex4-2}
\usepackage[utf8]{inputenc}
\usepackage[T1]{fontenc}
\usepackage{graphicx}
\usepackage[version=4]{mhchem}
\usepackage[hidelinks]{hyperref}
\usepackage{xcolor}
\usepackage{siunitx}
\usepackage{booktabs}
\usepackage[normalem]{ulem}
\usepackage{comment}

\newcommand{\FeTix}{\ensuremath{\mathrm{Fe}_{\mathrm{Ti}}^{\times}}}
\newcommand{\FeTip}{\ensuremath{\mathrm{Fe}_{\mathrm{Ti}}^{\prime}}}
\newcommand{\FeTipp}{\ensuremath{\mathrm{Fe}_{\mathrm{Ti}}^{\prime\prime}}}
\newcommand{\FeTi}{\ensuremath{\mathrm{Fe}_{\mathrm{Ti}}}}
\newcommand{\VOx}{\ensuremath{V_{\mathrm{O}}^{\times}}}
\newcommand{\VOp}{\ensuremath{V_{\mathrm{O}}^{\bullet}}}
\newcommand{\VOpp}{\ensuremath{V_{\mathrm{O}}^{\bullet\bullet}}}
\newcommand{\VO}{\ensuremath{V_{\mathrm{O}}}}

\newcommand{\stepO}{$\text{*}\mathrm{OH}\!\rightarrow\!\text{*}\mathrm{O}$}
\newcommand{\stepOOH}{$\text{*}\mathrm{O}\!\rightarrow\!\text{*}\mathrm{OOH}$}
\newcommand{\stepOtwo}{$\text{*}\mathrm{OOH}\!\rightarrow\!\text{*}\!+\mathrm{O}_2$}
\begin{document}

\title{Fermi-Level-Dependent Defect Chemistry and Oxygen Evolution Reaction Activity of Fe-Doped and Oxygen-Deficient \ce{SrTiO3}(001)}

\author{Amit Sehrawat}
\email{amit.sehrawat@tu-darmstadt.de}
\affiliation{Institut für Materialwissenschaft, Technische Universität Darmstadt, Otto-Berndt-Strasse 3, 64287 Darmstadt, Germany}
\author{Jochen Rohrer}
\affiliation{Institut für Materialwissenschaft, Technische Universität Darmstadt, Otto-Berndt-Strasse 3, 64287 Darmstadt, Germany}
\date{12.06.2026}

\begin{abstract}
The oxygen evolution reaction (OER) on perovskite oxides is controlled by
the interplay of dopant chemistry, defect charge states, and surface
segregation, yet these factors are rarely treated on equal footing. Using
first-principles density functional theory, we investigate how Fe dopants
(\FeTi{}) and oxygen vacancies (\VO{}) in different charge states affect
the OER on TiO$_2$-terminated \ce{SrTiO3}(001). We combine charge-dependent
defect formation energies, segregation energies, and charge transition
levels with OER free-energy profiles obtained in the computational hydrogen
electrode framework. Neutral \FeTix{} preserves near-pristine activity,
with overpotentials of $0.43$--$0.48$~V compared to $0.45$~V for the
pristine surface, whereas the reduced states \FeTip{} and \FeTipp{}
raise the overpotential to as much as $1.35$~V when intermediates bind to
Ti sites adjacent to surface Fe. Oxygen vacancies segregate to the surface
across the entire band gap ($\Delta E_\mathrm{seg} = -0.50$ to $-0.80$~eV)
but do not improve the activity: \VOx{} and \VOp{} overstabilize oxygenated
intermediates ($\eta$ up to $2.13$~V), and only \VOpp{} retains a balanced
pathway ($\eta = 0.45$~V in the bulk-like region). Because the stable
charge state and the segregation tendency of each defect are set by the
Fermi level, the OER overpotential itself becomes a Fermi-level-dependent
quantity. These results establish Fermi-level engineering as a framework
for assessing and tuning defect-mediated OER activity in perovskite
oxides.
\end{abstract}

\maketitle

\section{Introduction}
Electrochemical and photoelectrochemical water splitting are promising
routes to green hydrogen, a clean energy carrier that can reduce dependence
on fossil fuels~\cite{Lewis2006, Seh2017, li2019recent,
raveendran2023comprehensive}. The overall efficiency of water electrolysis
is, however, limited by the sluggish kinetics of the oxygen evolution
reaction (OER) at the anode, which proceeds through four coupled
proton--electron transfer steps and several oxygenated
intermediates~\cite{Rossmeisl2007, man2011universality}. The design of active, stable,
and inexpensive OER electrocatalysts therefore remains a central challenge
in electrochemistry~\cite{xie2022oxygen, Suen2017}.

Perovskite oxides of the form ABO$_3$ are attractive OER catalysts because
their activity can be tuned through cation substitution, oxygen
nonstoichiometry, and surface defect chemistry~\cite{liu2024perovskite,
Hwang2017}. The electronic structure of the B-site cation, its oxidation
state, and the covalency with lattice oxygen control the binding of the
*OH, *O, and *OOH intermediates and underlying mechanism thereby the
overpotential~\cite{Suntivich2011, calle2013number, grimaud2017activating, Man2011}.



\ce{SrTiO3} (STO) is among the most widely investigated perovskite oxides, owing to its chemical stability, well-defined surfaces, and long history in photocatalysis~\cite{wrighton1976strontium, domen1980photocatalytic, avcioǧluphotocatalytic, zhang2024crystal}. Pristine \ce{SrTiO3} is nevertheless a poor dark-OER catalyst because of its wide band gap, low
electronic and ionic conductivity, and the absence of redox-active transition-metal
states near the Fermi level~\cite{Akbashev2018}. Several doping strategies
address these limitations. Akbashev \textit{et al.}~\cite{Akbashev2018}
showed that a single subsurface SrRuO$_3$ layer activates the \ce{SrTiO3}
surface for the OER. Al doping suppresses sub-band-gap states and reduces
electron--hole recombination at Ti sites~\cite{zhao2019electronic}, and
Al/La co-doping lowers the oxygen-vacancy concentration and enhances
photocatalytic overall water splitting~\cite{qin2021codoped}.

Among B-site dopants, Fe is of particular interest because it adopts
multiple oxidation states in the
SrTi$_{1-x}$Fe$_x$O$_{3-\delta}$ (STF) lattice, with the Fe valence
directly coupled to the oxygen nonstoichiometry and the electronic
conductivity~\cite{rothschild2006electronic, kubacki2018impact,
merkle2008how}. Fe doping introduces occupied $3d$ states above the valence
band maximum of \ce{SrTiO3}, narrowing the band gap and enhancing
visible-light absorption~\cite{zhou2011effect, rothschild2006electronic} and catalytic activity.
Experimentally, the OER activity of STF electrodes increases with Fe
content: Hayden and Rogers reported a monotonic decrease of the OER onset
potential with increasing $x$ in compositionally graded STF
films~\cite{hayden2018oxygen}, Lankauf \textit{et al.} observed improved
OER performance at higher Fe content in STF
powders~\cite{lankauf2021effect}, and STF electrodes operate stably under
oxygen-evolution conditions in solid oxide cells~\cite{zhang2019high}.

Despite these observations, the atomic-scale role of Fe in \ce{SrTiO3}
remains unclear, because Fe can influence the OER in several coupled ways.
First, Fe can occupy surface or bulk-like Ti sites, and its segregation
tendency determines whether it acts directly as an adsorption site or
modifies nearby Ti sites from a subsurface position. Second, the charge
state of the substitutional defect \FeTi{} changes the binding strength of
the OER intermediates. In Kr\"oger--Vink notation, which we use throughout,
the charge states $q = 0$, $-1$, and $-2$ of \FeTi{} are written \FeTix{},
\FeTip{}, and \FeTipp{} and correspond formally to Fe$^{4+}$, Fe$^{3+}$,
and Fe$^{2+}$ substituting for Ti$^{4+}$; likewise, the oxygen-vacancy
charge states $q = 0$, $+1$, and $+2$ are written \VOx{}, \VOp{}, and
\VOpp{}. Both the spatial distribution of Fe and its oxidation state must
therefore be treated on equal footing when connecting Fe doping to OER
activity.

Oxygen vacancies add a second, coupled degree of freedom. They modify the
electronic and surface structure of \ce{SrTiO3} and can determine whether the
OER proceeds through the adsorbate evolution mechanism or involves lattice
oxygen~\cite{Mefford2016, Grimaud2017, Yoo2018}. For TiO$_2$-terminated
\ce{SrTiO3}(001), neutral oxygen vacancies at both surface and bulk-like
sites were shown to increase the OER
overpotential~\cite{cui2019oxygen, sokolov2024computational} in first-principles calculations. However, the
charge state of the vacancy was not resolved in these studies. It thus
remains open whether oxygen vacancies in different charge states segregate
to the surface or remain in bulk-like regions, and how vacancy charge state
and position jointly affect the OER.

Here, we use first-principles calculations to investigate how Fe dopant
position, Fe charge state, and oxygen vacancies control the OER on
TiO$_2$-terminated \ce{SrTiO3}(001). We compute charge-dependent defect
formation energies, segregation energies, and charge transition levels
within the established point-defect formalism~\cite{freysoldt2014first},
and combine them with OER free-energy profiles obtained in the
computational hydrogen electrode (CHE)
framework~\cite{Norskov2004, Rossmeisl2007}. We find that the catalytic
behavior is governed by a strong coupling between Fermi-level-dependent
defect stability and intermediate binding: neutral \FeTix{} preserves
near-pristine activity ($\eta = 0.43$--$0.48$~V vs.\ $0.45$~V for the
pristine surface), whereas the reduced states \FeTip{} and \FeTipp{} and
surface oxygen vacancies destabilize the adsorbate evolution pathway and
raise the overpotential by up to $\sim$1.7~V. This establishes a direct
link between defect thermodynamics, Fermi-level position, and surface
catalytic activity in Fe-doped \ce{SrTiO3}.

\section{Methods}
\label{sec:methods}
\subsection{Computational details}
All calculations were performed using density functional theory as
implemented in the Vienna Ab initio Simulation Package
(VASP)~\cite{kresse1996efficient, kresse1996efficiency}. We used the PBEsol functional within the generalized gradient approximation (GGA)~\cite{perdew1996generalized}, as it offers a practical balance between computational efficiency and accuracy for solid oxides.
The projector augmented wave (PAW) method~\cite{blochl1994projector,kresse1999ultrasoft} was
used with potentials from the VASP library, treating the Sr
$4s^2 4p^6 5s^2$ (Sr\_sv), Ti $3p^6 3d^3 4s^1$ (Ti\_pv),
O $2s^2 2p^4$, Fe $3d^6 4s^2$, and H $1s^1$ electrons as valence states.
The plane wave kinetic energy cutoff was set to \SI{520}{eV}. A Gaussian
smearing scheme with a width of \SI{0.1}{eV} was applied. Electronic self
consistency was converged to \SI{1e-5}{eV}. For bulk primitive cell
calculations, Brillouin zone sampling employed a
$9 \times 9 \times 9$ $\Gamma$ centered Monkhorst–Pack~\cite{monkhorst1976special} $k$ point mesh, and
ionic relaxations were continued until all forces were below
\SI{0.01}{eV/\angstrom}. For slab calculations, a
$3 \times 3 \times 1$ $\Gamma$ centered Monkhorst–Pack $k$ point mesh was
used, and the structures were relaxed until all forces were below
\SI{0.03}{eV/\angstrom}. Spin-polarized calculations were carried out for all systems.

\subsection{Defect formation energies and segregation energies}
\label{sec:defect_formation_segregation}

The formation energy of a defect $D$ in charge state $q$ was evaluated
within the standard formalism~\cite{freysoldt2014first} as
\begin{equation}
    \begin{split}
        E_\mathrm{form}(D^q) = {} & E_\mathrm{tot}(D^q) - E_\mathrm{tot}(\mathrm{host}) \\
        & - \sum_i n_i \mu_i + q(E_\mathrm{F} + E_\mathrm{VBM}) + E_\mathrm{corr},
    \end{split}
    \label{eq:defect_formation}
\end{equation}
where $E_\mathrm{tot}(D^q)$ and $E_\mathrm{tot}(\mathrm{host})$ are the
total energies of the defective and pristine cells, $n_i$ is the number of
atoms of species $i$ added to ($n_i>0$) or removed from ($n_i<0$) the
supercell, and $\mu_i$ are the corresponding chemical potentials. The Fermi
level $E_\mathrm{F}$ is referenced to the valence band maximum
$E_\mathrm{VBM}$, and $E_\mathrm{corr}$ is the finite-size electrostatic
correction for charged defects, computed with the FNV scheme and its
repeated-slab extension as implemented in
\textit{sxdefectalign2d}~\cite{freysoldt2009fully, freysoldt2018first}.

Chemical potentials were fixed at the O-rich limit
($\Delta\mu_\mathrm{O}=0$~eV), with the oxygen chemical potential
referenced to the \ce{O2} molecule and corrected for the well-known
GGA overbinding of \ce{O2}~\cite{wang2006oxidation} by
$0.687$~eV per O atom, following the MP2020
scheme~\cite{wang2021framework}, i.e.,
$\mu_{\ce{O}} = \tfrac{1}{2}\epsilon_{\ce{O2}} - 0.687~\mathrm{eV}
= -4.440$~eV.
The remaining chemical potentials follow from the equilibrium conditions
$\mu_{\ce{Ti}} = \epsilon_{\ce{TiO2}} - 2\mu_{\ce{O}}$ (from
\ce{TiO2} equilibrium) and
$\mu_{\ce{Sr}} = \epsilon_{\ce{SrTiO3}} - \mu_{\ce{Ti}} - 3\mu_{\ce{O}}$
(from \ce{SrTiO3} equilibrium). For Fe, \ce{Fe2O3} served as the
reference, $\mu_{\ce{Fe}} = \tfrac{1}{2}(\epsilon_{\ce{Fe2O3}} - 3\mu_{\ce{O}})$.

From the formation energies of defects in the surface layer,
$E_\mathrm{form}(D_\mathrm{surf}^q)$, and in the bulk-like central region
of the slab, $E_\mathrm{form}(D_\mathrm{bulk}^q)$, we define the
segregation energy
\begin{equation}
  \Delta E_\mathrm{seg}(D^q)
    = E_\mathrm{form}(D_\mathrm{surf}^q)
    - E_\mathrm{form}(D_\mathrm{bulk}^q),
  \label{eq:seg}
\end{equation}
such that $\Delta E_\mathrm{seg}<0$ indicates thermodynamic stabilization
at the surface and $\Delta E_\mathrm{seg}>0$ a preference for the bulk-like
position. Thermodynamic charge transition levels were obtained as
\begin{equation}
    \varepsilon(q/q') =
    \frac{
    E_\mathrm{form}(D^q;E_\mathrm{F}=0)
    -
    E_\mathrm{form}(D^{q'};E_\mathrm{F}=0)
    }{q'-q},
    \label{eq:ctl}
\end{equation}
where $E_\mathrm{F}=0$ corresponds to the valence band maximum.

For \FeTi{}, the charge states $q=0$, $-1$, and $-2$ (\FeTix{}, \FeTip{},
\FeTipp{}) were considered; for \VO{}, the charge states $q=0$, $+1$, and
$+2$ (\VOx{}, \VOp{}, \VOpp{}).

\subsection{OER calculations}
\label{sec:oer_methods}

The OER was studied within the computational hydrogen
electrode (CHE) framework \cite{Norskov2004,Rossmeisl2007}. 
In this approach,
the free energy of the proton and electron pair (H$^+$ + $e^-$) is referenced
to $\frac{1}{2}G(\text{H}_2)$ at standard conditions, and the effect of an
applied potential $U$ enters as $-eU$ per electron transferred. The four
elementary steps are:
\begin{subequations}
\label{eq:oer}
\begin{align}
  * + \mathrm{H_2O\,(l)}
    &\longrightarrow \mathrm{{}^{*}OH} + \mathrm{H^+} + e^-,
    & \Delta G_1 \label{eq:step1} \\
  \mathrm{{}^{*}OH}
    &\longrightarrow \mathrm{{}^{*}O} + \mathrm{H^+} + e^-,
    & \Delta G_2 \label{eq:step2} \\
  \mathrm{{}^{*}O} + \mathrm{H_2O\,(l)}
    &\longrightarrow \mathrm{{}^{*}OOH} + \mathrm{H^+} + e^-,
    & \Delta G_3 \label{eq:step3} \\
  \mathrm{{}^{*}OOH}
    &\longrightarrow * + \mathrm{O_2\,(g)} + \mathrm{H^+} + e^-,
    & \Delta G_4 \label{eq:step4}
\end{align}
\end{subequations}
The theoretical overpotential is
$\eta = \max_i(\Delta G_i)/e - \SI{1.23}{V}$, and the step with the largest
$\Delta G_i$ is referred to as the potential limiting step (PLS) in the
thermodynamic sense. For each intermediate (*OH, *O, *OOH), multiple
adsorption geometries were sampled on the active sites
(Sec.~\ref{sec:oer_pristine}), and the lowest-energy geometry was used in
the free-energy analysis.

Free energies were obtained as
$\Delta G = \Delta E_\mathrm{DFT} + \Delta E_\mathrm{ZPE} - T\Delta S$,
with zero-point energy and entropy contributions for the adsorbates and
gas-phase molecules taken from Ref.~\cite{wen2021strain}
(Table~\ref{tab:zpe_ts_corrections}); these were computed for the pristine
TiO$_2$-terminated SrTiO$_3$(001) surface with the PBE functional. Because
such corrections depend on substrate, termination, functional, and
adsorbate geometry---and, for defective surfaces, additionally on defect
location and charge state---we apply the same correction set to all systems
to compare trends consistently. 

\begin{table}[tb]
\centering
\caption{Free energy corrections, \(\mathrm{ZPE}-TS\), used for adsorbed OER intermediates and gas phase molecules. The values are taken from Ref.~\cite{wen2021strain}. For \(\mathrm{H_2O}\)(g), the correction corresponds to \(0.035~\mathrm{bar}\) and \(298.15~\mathrm{K}\). Energies are given in eV.}
\label{tab:zpe_ts_corrections}
\begin{tabular}{lc}
\toprule
Species & \(\mathrm{ZPE} - TS\) \\
\midrule
\(*\mathrm{OH}\) & 0.275 \\
\(*\mathrm{O}\) & -0.014 \\
\(*\mathrm{OOH}\) & 0.337 \\
\(\mathrm{H_2}\)(g) & -0.14 \\
\(\mathrm{H_2O}\)(g) & -0.11 \\
\bottomrule
\end{tabular}
\end{table}

\section{Results and Discussion}

In order to examine the dependence of OER overpotentials on the Fermi level we first
compute overpotentials for the pristine STO surface (Sec.~\ref{sec:oer_pristine}).
We then investigate defect formation energies and surface segregation trends for oxygen vacancies and iron dopants in various oxidation states (Sec.~\ref{sec:defect_formation}).
subsequently, we calculate overpotentials in the presence of these defects 
(Sec.~\ref{sec:oer_defects})
and eventually relate these overpotentials to the Fermi level (Sec.~\ref{sec:fermi_discussion}).

\subsection{OER activity: Pristine surfaces}
\label{sec:oer_pristine}

\begin{figure}[tb]
    \centering
    \includegraphics[width=0.95\linewidth]{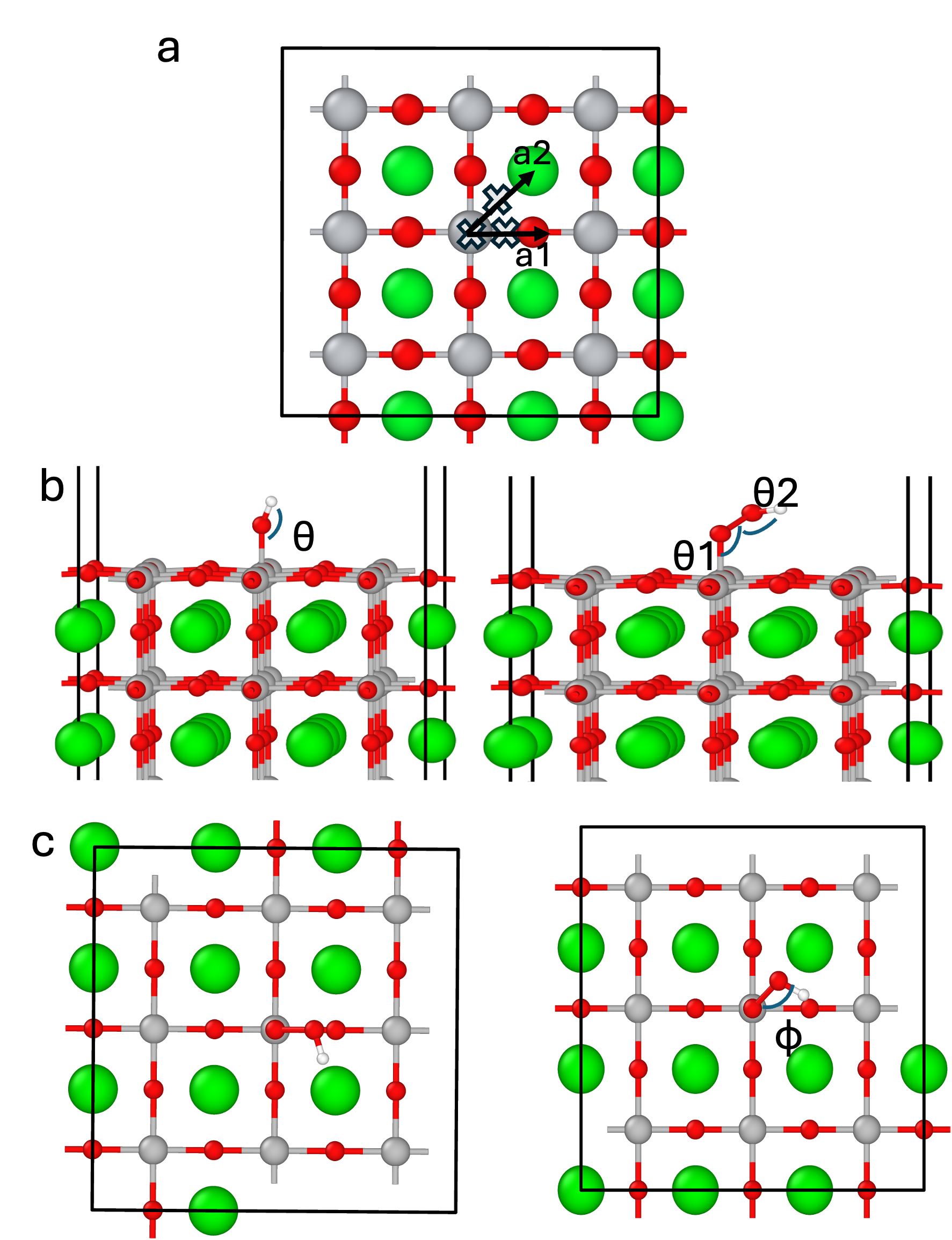}
    \caption{Initial adsorption configurations on pristine
    TiO$_2$-terminated \ce{SrTiO3}(001).
    (a) Top view of the surface with the three high-symmetry adsorption
    sites (white crosses): Ti top, Ti--lattice-O bridge, and
    Ti--subsurface-Sr bridge.
    (b) Sampled orientations of adsorbed *OH, defined by the Ti--O--H
    angle $\theta$, and of *OOH,
    defined by the fixed Ti--O--O angle $\theta_1 = 135^\circ$ and the
    O--O--H angle
    $\theta_2$.
    (c) In-plane alignment of each adsorbate along the $a_1$ or $a_2$
    surface direction. Color coding: green = Sr, grey = Ti, red = O, white = H.}
    \label{fig:Initial_configs}
\end{figure}

The pristine TiO$_2$-terminated \ce{SrTiO3}(001) surface is shown in Fig.~\ref{fig:Initial_configs} (1a). Three high-symmetry adsorption sites for the OER intermediates *O, *OH, and *OOH were considered: on top of surface Ti, at the Ti--lattice O bridge, and at the Ti--subsurface Sr bridge shown by crosses in Fig.~\ref{fig:Initial_configs} (1a).
For OH and OOH, additional orientational degrees of freedom were included. For OH, the Ti--O--H angle, $\theta$, was varied between $45^\circ$, and $180^\circ$ as shown in Fig.~\ref{fig:Initial_configs} (b). For OOH, two angles were
considered: the Ti--O--O angle, $\theta_1$, and the O--O--H angle,
$\theta_2$. The angle $\theta_1$ was fixed at $135^\circ$, while
$\theta_2$ was varied in steps, rotating the H atom from pointing
toward the surface to pointing away from it.
As a second degree of freedom, each fixed molecular geometry was aligned along either the $a_1$ or $a_2$ surface direction, as shown in Fig.~\ref{fig:Initial_configs} (c). This corresponds to a rigid rotation of the adsorbate around the surface normal in the $xy$ plane.

Upon relaxation, we find that various initial states yield (almost) identical final configurations.
Representative optimized adsorption geometries are provided in the
Supplementary Material (SM; Figs.~S1 and~S2).
Fig.~\ref{fig:pristine_oer_profile} shows the free energy profile of the OER
intermediates on the \ce{TiO2} terminated STO(001) surface 
obtained using the minimum energy configurations identified in our optimization procedure. 
We identify
$^*\text{OH} \rightarrow {^*\text{O}}$ as the PLS. The
minimum energy configuration of the adsorbed $^*$O species corresponds to an O
atom bridging the top Ti site and a neighboring lattice O site, yielding an
overpotential of $\eta = 0.45$~V. For the alternative dangling configuration,
where the adsorbed O atom sits directly atop the Ti site, we obtain a higher
overpotential of $\eta = 1.16$~V. This configuration is approximately
0.71~eV higher in energy than the bridge site. The relative stability and
overpotential ordering of the two sites are consistent with previous reports,
which also found the bridge site to be more favorable and associated with a
substantially lower overpotential than the dangling site~\cite{cui2019oxygen, sokolov2024computational}.

\begin{figure}[tb]
    \centering
    \includegraphics[width=0.95\columnwidth]{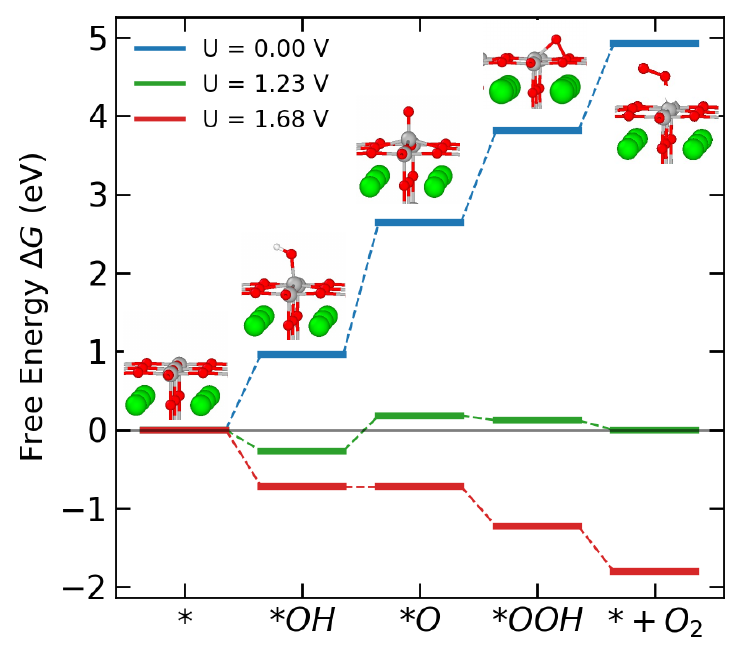}
    \caption{
    Free energy diagram for the OER on pristine TiO$_2$ terminated
    \ce{SrTiO3}(001). The OER intermediates are shown along the reaction
    coordinate as *, *OH, *O, *OOH, and * + O$_2$. At $U=0$ V,
    the largest uphill step is approximately $1.68$ eV, corresponding to a
    theoretical overpotential of $\eta = 0.45$ V. At $U=1.23$ V, the profile
    remains limited by the most uphill oxygen intermediate formation step,
    whereas at $U=1.68$ V the thermodynamic OER cycle becomes downhill.
    }
    \label{fig:pristine_oer_profile}
\end{figure}

Our calculated overpotentials differ moderately from earlier values for the
bridge site, $\eta = 0.71$~V~\cite{cui2019oxygen} and
$0.65$~V~\cite{sokolov2024computational}, and agree more closely for the
dangling site, where reported values are $\eta = 1.26$~V~\cite{cui2019oxygen}
and $1.14$~V~\cite{sokolov2024computational}. These differences likely arise
from the combined effect of exchange and correlation functional, surface
geometry, slab symmetry, in-plane cell size, and spin treatment. In
particular, Ref.~\cite{sokolov2024computational} employed an $R\bar{3}c$
\ce{SrTiO3} structure, whereas Ref.~\cite{cui2019oxygen} used a
$2\times2$ surface cell with asymmetric termination. Both studies used the PBE
functional, while the present work uses PBEsol. Despite the sensitivity of the absolute overpotential to the computational setup, the consistent site ordering preserves the qualitative OER pathway.


\begin{figure}[tb]
    \centering
    \includegraphics[width=0.95\linewidth]{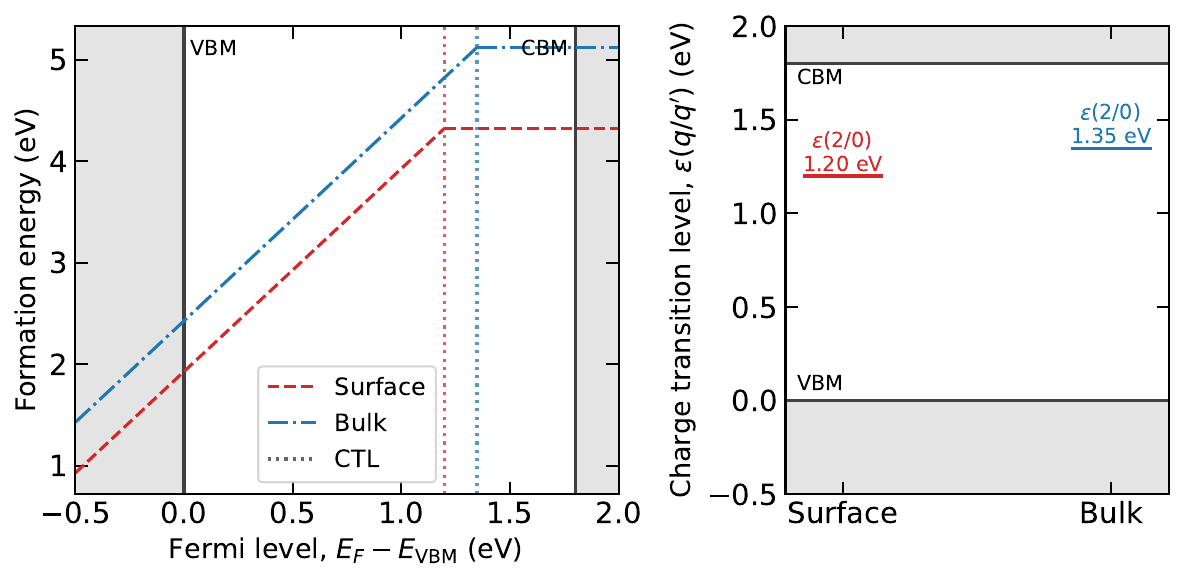}\\
    \includegraphics[width=0.95\linewidth]{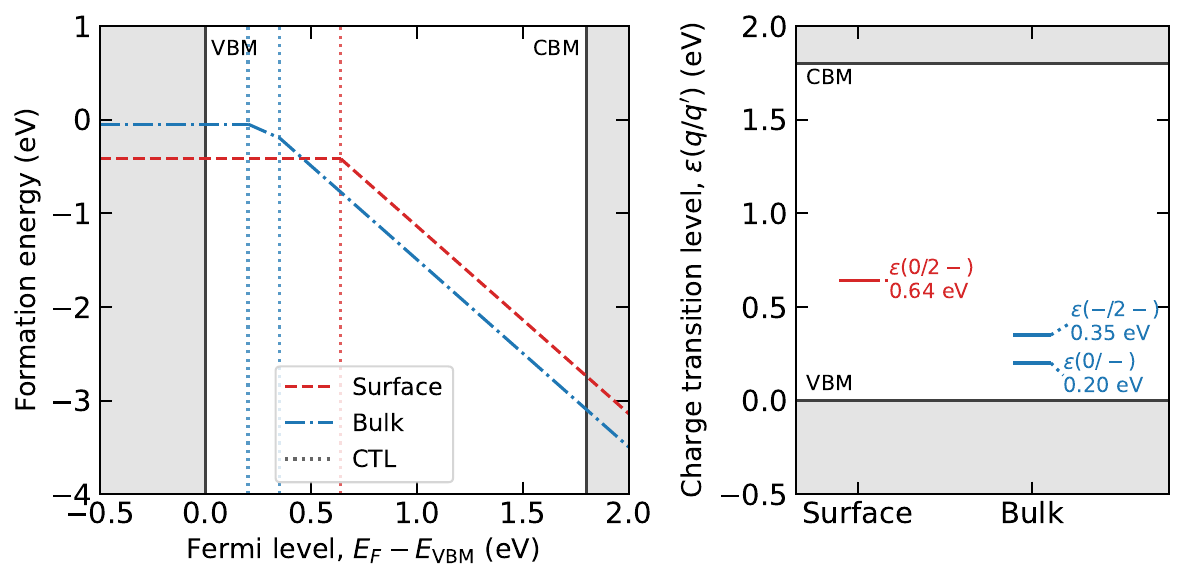}
    \caption{Defect formation energies as a function of the Fermi level
    $E_\mathrm{F}$ (referenced to the VBM) for
    (top) oxygen vacancies \VO{} and (bottom) Fe dopants \FeTi{} at
    surface and bulk-like sites of
    TiO$_2$-terminated \ce{SrTiO3}(001). Line slopes correspond to the
    defect charge state $q$; kinks mark the charge transition levels
    $\varepsilon(q/q')$ of Eq.~\eqref{eq:ctl}. For \VO{},
    $\varepsilon(2/0) = 1.20$~eV (surface) and $1.35$~eV (bulk-like);
    \VOp{} never appears on the lowest-energy envelope. For \FeTi{}, the
    bulk-like defect shows $\varepsilon(0/-1)=0.20$~eV and
    $\varepsilon(-1/-2)=0.35$~eV, whereas the surface defect shows a
    direct $\varepsilon(0/-2)=0.64$~eV transition. The vertical offset
    between surface and bulk-like branches gives the segregation energy of
    Eq.~\eqref{eq:seg}.}
    \label{fig:defect_formation_ctls}
\end{figure}

\subsection{Defect formation energies, charge transition levels, and segregation tendencies}
\label{sec:defect_formation}

In Fig.~\ref{fig:defect_formation_ctls}, we present the calculated defect formation energies and the resulting charge-transition levels (CTLs) for oxygen vacancies ($V_\mathrm{O}$, top panel) and iron dopants (Fe$_\mathrm{Ti}$, bottom panel). In both cases, the defect formation energies depend not only on the Fermi level, but also on the location of the defect.

Oxygen vacancies can be stabilized either as doubly positively charged vacancies, $V_{\mathrm{O}}^{\bullet\bullet}$, or as neutral vacancies, $V_{\mathrm{O}}^{\times}$, while $V_{\mathrm{O}}^{\bullet}$ is never thermodynamically stabilized. The CTLs are located at $\varepsilon(2/0)=1.20$~eV and $1.35$~eV for the surface and bulk-like vacancies, respectively. Independent of the Fermi level, $V_\mathrm{O}$ generally tends to segregate to the surface, if kinetically possible, with segregation energies of $\Delta E_\mathrm{seg}(V_{\mathrm{O}}^{\bullet\bullet}) = -0.50$~eV 
and $\Delta E_\mathrm{seg}(V_{\mathrm{O}}^{\times}) = -0.80$~eV.

For Fe$_\mathrm{Ti}$, the situation is different. The bulk-like defect has two transition levels, namely $\varepsilon(0/-1)=0.20$~eV above the VBM and $\varepsilon(-1/-2)=0.35$~eV, such that Fe$_\mathrm{Ti}$ may exist as Fe$_\mathrm{Ti}^{\times}$, Fe$_\mathrm{Ti}^{'}$, and Fe$_\mathrm{Ti}^{''}$. At the surface, Fe$_\mathrm{Ti}$ exists only as Fe$_\mathrm{Ti}^{\times}$ or Fe$_\mathrm{Ti}^{''}$, with a direct transition level $\varepsilon(0/-2)=0.64$~eV above the VBM. Moreover, surface segregation of Fe$_\mathrm{Ti}$ is favored only as Fe$_\mathrm{Ti}^{\times}$ and only up to a Fermi level of approximately $0.46$~eV. For higher values of $E_F$, the bulk-like Fe$_\mathrm{Ti}^{''}$ state is favored.

The trend identified for $V_\mathrm{O}$ is consistent with the recent defect formation energy study of perovskite surfaces by Ned \textit{et al.}~\cite{w683-tvc4}, where surface oxygen vacancies were found to be stabilized for the majority of titanate surfaces, including \ce{TiO2}-terminated \ce{SrTiO3}. In particular, they reported that the neutral surface oxygen vacancy is approximately $1$~eV more favorable than the corresponding bulk-like vacancy. Their results for \ce{SrSnO3} also show a direct $\varepsilon(2/0)$ transition, with $V_{\mathrm{O}}^{\times}$ stable over a large part of the band gap, and a shift of the $V_{\mathrm{O}}^{\bullet\bullet}/V_{\mathrm{O}}^{\times}$ transition to higher Fermi levels for more bulk-like vacancy positions. Our results therefore follow the same general behavior: oxygen vacancies are thermodynamically stabilized at the surface, and moving the vacancy into a more bulk-like environment shifts the charge-transition level upward in energy.

\subsection{OER activity: Defective surfaces}
\label{sec:oer_defects}
We next investigate the OER free-energy landscape and corresponding overpotentials in the presence of surface and bulk defects. 
For bulk defects, we consider all metastable structures found for adsorption on the pristine surface, add the bulk defect and perform further relaxations in the various charge states. 
For the surface defects additional initial configurations were examined since the defect breaks the symmetry.
Corresponding initial states are shown in Fig. S3 in the SM.
Overpotentials are again computed from the minimum-energy configurations.
 
For the Fe-doped systems, three configurations were evaluated for the
OER calculations: Fe substituting a surface Ti site with the adsorbate
on the Fe site; Fe substituting a surface Ti site with the adsorbate on
a neighboring Ti site; and Fe substituting a bulk-like Ti site with the
adsorbate on a surface Ti site. For the oxygen-vacancy systems, two
configurations were considered: $V_\mathrm{O}$ located at the surface
with the adsorbate on a neighboring Ti site; and $V_\mathrm{O}$ located
in a bulk-like region with the adsorbate on a surface Ti site.

\begin{figure*}[tb]
    \centering
    \includegraphics[width=0.9\textwidth]{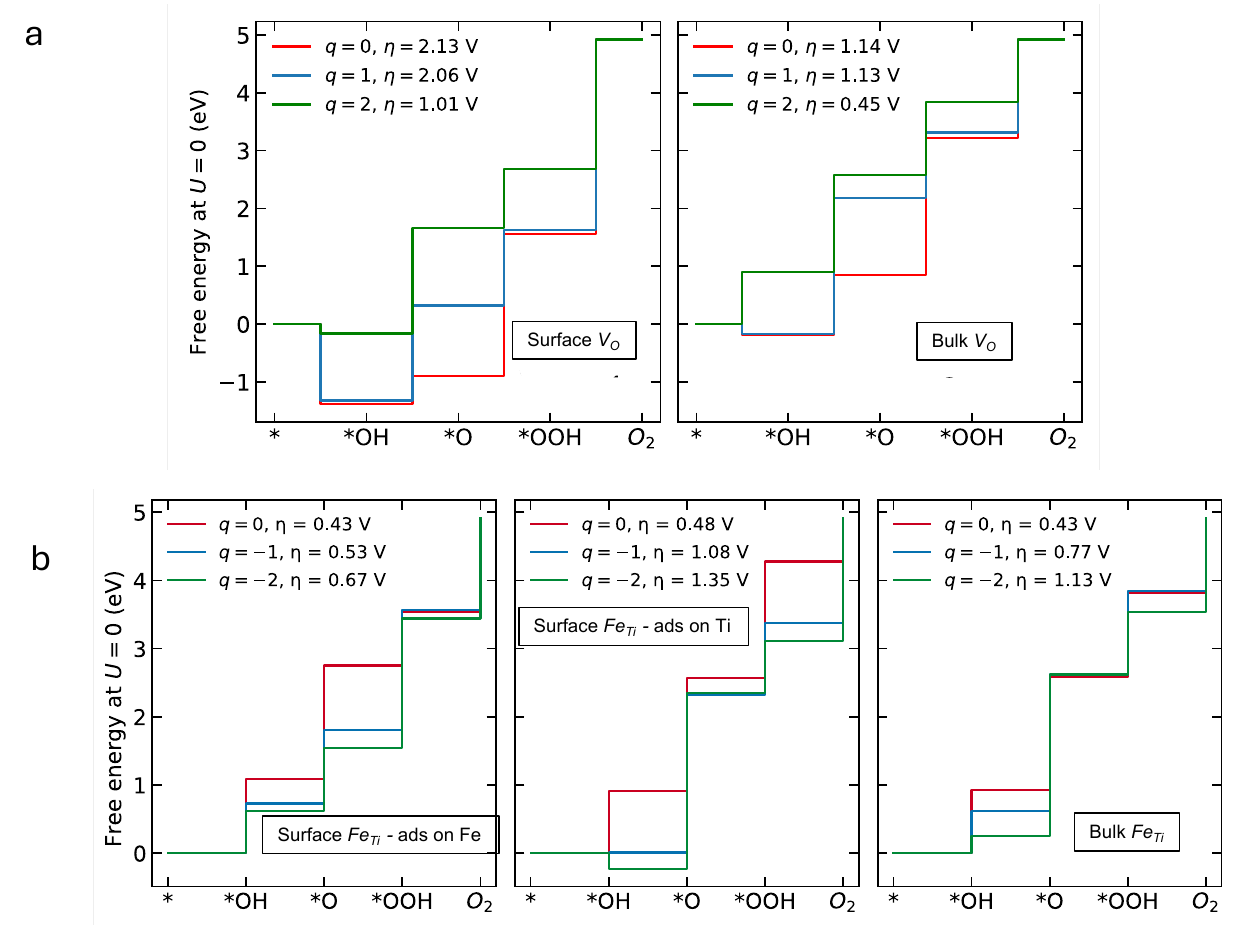}
    \caption{OER free energy profiles for (a) $V_\mathrm{O}$ (b) $\mathrm{Fe_{Ti}}$ for \ce{TiO2}-terminated \ce{SrTiO3}(001) surface as a function of charge states and adsorption sites.}
    \label{fig:combine_oer_profiles}
\end{figure*}

\begin{figure}[tb]
    \centering
    \includegraphics[width=\columnwidth]{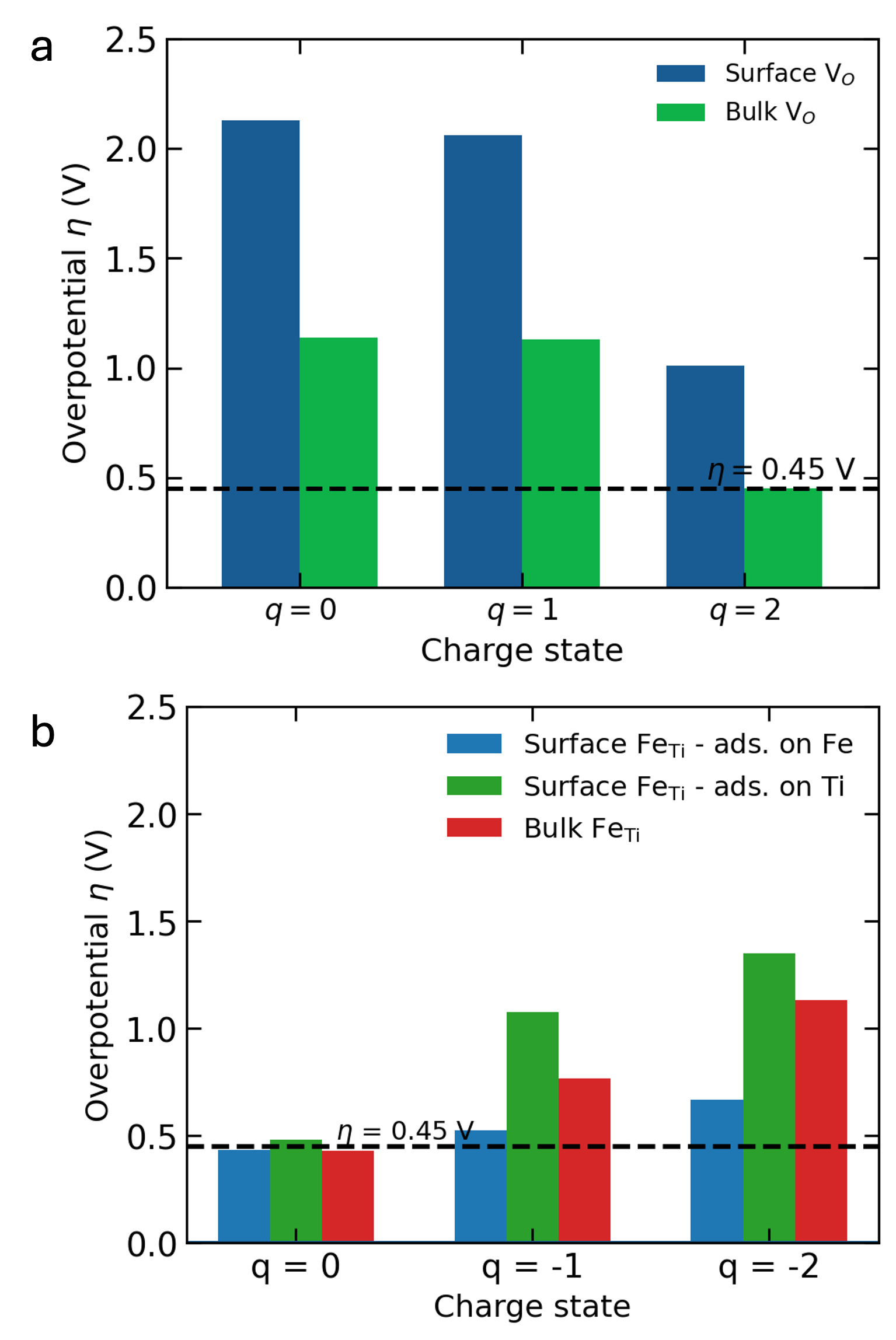}
    \caption{Calculated OER overpotentials for (a) $V_\mathrm{O}$ and (b) $\mathrm{Fe_{Ti}}$ defects for \ce{TiO2}-terminated \ce{SrTiO3}(001) surface as a function of charge state and adsorption site. The black dashed horizontal line marks the
    pristine surface overpotential of $\eta = 0.45$ V.
    }
    \label{fig:combine_eta}
\end{figure}


\subsubsection{Oxygen vacancies}
For oxygen vacancies, the OER free-energy profiles and overpotentials are shown in the 
top panels of Figs.~\ref{fig:combine_oer_profiles} and~\ref{fig:combine_eta}.

For the bulk-like oxygen vacancy, the calculated overpotentials are
$\eta = 1.14$~V for $V_{\mathrm{O}}^{\times}$, $\eta = 1.13$~V for
$V_{\mathrm{O}}^{\bullet}$, and $\eta = 0.45$~V for
$V_{\mathrm{O}}^{\bullet\bullet}$. Only the fully ionized
$V_{\mathrm{O}}^{\bullet\bullet}$ strongly improves the OER free-energy
profile, reducing the overpotential to $\eta = 0.45$~V, while the
neutral and singly charged states leave it high at above $1.1$~V.
For $V_{\mathrm{O}}^{\times}$, the limiting step is the formation of *OOH from *O,
with $\Delta G_3=2.37$ eV. For $V_{\mathrm{O}}^{\bullet}$, the limiting step shifts
to the oxidation of *OH to *O, with $\Delta G_2=2.36$ eV. In contrast,
$V_{\mathrm{O}}^{\bullet\bullet}$ gives a much more balanced profile, with the largest step
being $\Delta G_2=1.68$ eV, corresponding to an overpotential of
$\eta=0.45$ V. This value is identical to what we obtained for the pristine
\ce{SrTiO3}(001) overpotential ($\eta=0.45$ V), indicating that a bulk-like
doubly charged oxygen vacancy does not degrade the adsorbate
evolution mechanism.

For surface oxygen vacancies, we observe the same trend as in the bulk,
where increasing the charge state of $V_\mathrm{O}$ improves the OER
overpotential, although the actual values differ substantially.
For surface $V_{\mathrm{O}}^{\times}$ and $V_{\mathrm{O}}^{\bullet}$ 
the overpotentials are $\eta=2.13$ V and $\eta=2.06$ V, respectively. In both
cases, the first proton and electron transfer step is strongly downhill
($\Delta G_1=-1.38$ eV for $q=0$ and $-1.32$ eV for $q=+1$), showing that
the *OH intermediate is overstabilized at the defective surface.
Consequently, the final \stepOtwo{} step becomes highly
uphill and determines the overpotential, with $\Delta G_4=3.36$ eV for
$V_{\mathrm{O}}^{\times}$ and $\Delta G_4=3.29$ eV for $V_{\mathrm{O}}^{\bullet}$. The
surface $V_{\mathrm{O}}^{\bullet\bullet}$ state again gives the most favorable OER energy-profile, with an overpotential of $\eta=1.01$ V. However, even this case
remains less active than the pristine surface and less favorable than the
bulk-like $V_{\mathrm{O}}^{\bullet\bullet}$ case. This is in qualitative agreement with a previous study by Cui et al.~\cite{cui2019oxygen} on the \ce{TiO2} terminated \ce{SrTiO3}(001) surface, which reported that neutral oxygen vacancies increase the overpotential.

These results show that both the charge state and the position of the oxygen
vacancy control the OER activity. The charge state is particularly important:
for both bulk-like and surface vacancies, $V_{\mathrm{O}}^{\bullet\bullet}$ gives the lowest
overpotential, while $V_{\mathrm{O}}^{\times}$ and $V_{\mathrm{O}}^{\bullet}$ lead to much
less favorable OER profiles. The position of the vacancy is also important,
especially for the neutral and singly charged states. Moving the vacancy from
the bulk-like region to the surface increases the overpotential from
$1.14$ to $2.13$ V for $q=0$ and from $1.13$ to $2.06$ V for $q=+1$.
Therefore, surface oxygen vacancies are particularly detrimental when they are neutral or singly charged.

\subsubsection{Fe dopants}
For iron dopants, the OER free-energy profiles and overpotentials are shown in
the bottom panels of Figs.~\ref{fig:combine_oer_profiles} and~\ref{fig:combine_eta}.

When Fe sits in bulk-like environment, the neutral state gives a low
$\eta = 0.43$~V with \stepO{} as the PLS, slightly below the pristine surface.
The charged states raise the overpotential to $0.77$~V for $q=-1$ and $1.13$~V
for $q=-2$, with \stepO{} remaining the PLS in all three charge states.
Subsurface Fe therefore acts indirectly, through
an electronic modification of the surface Ti adsorption site, with an
effect that is small for \FeTix{} but grows as negative charge is added.

When the adsorbate binds directly to the surface Fe, the neutral
state gives $\eta = 0.43$~V with \stepO{} as the PLS, comparable to the pristine
surface. Adding negative charge raises the overpotential to $0.53$~V for $q=-1$
and $0.67$~V for $q=-2$, and shifts the PLS from \stepO{} to \stepOOH{}. The
negative charge makes formation of the *OOH intermediate less favorable relative
to the preceding *O state.

When the surface Fe is retained but the adsorbate binds to a neighboring Ti site, the neutral state gives $\eta = 0.48$~V with \stepOOH{} as the PLS,
again close to pristine. The charged states show the largest overpotentials of
any Fe configuration: $1.08$~V for $q=-1$ and $1.35$~V for $q=-2$, with \stepO{}
as the PLS in both. The large $\Delta G_2$ values in
Fig.~\ref{fig:combine_oer_profiles}(b, mid panel) show that the *OH state is stabilized too
strongly relative to *O, making the deprotonation and oxidation from *OH to *O
the thermodynamic bottleneck. Placing the adsorbate on a Ti site adjacent to a
surface Fe dopant is therefore strongly unfavorable for negatively charged
defects.

Overall, the Fe activity is governed by a coupled dependence on dopant position,
adsorbate binding site, and charge state. The neutral $\mathrm{Fe_{Ti}}$ systems
give overpotentials of $0.43$--$0.48$~V across all sites, comparable to the
pristine case. Negative charge raises the overpotential in every configuration,
most severely when the adsorbate binds to a Ti site next to a surface Fe dopant.
The most unfavorable case is when Fe sits at surface and adsorbate binds to nearby Ti with charge state $q=-2$, where $\eta = 1.35$~V and the
PLS is \stepO{}. The most favorable Fe cases are the neutral
bulk Fe and surface Fe dopants, both results in overpotential of $0.43$~V, even though Gibbs free energies of adsorbates differ for both case resulting overpotential dictated by PLS \stepO.

\subsection{Fermi-level dependence of the OER overpotential}
\label{sec:fermi_discussion}
\begin{figure}[htb!]
    \centering
    \includegraphics[width=\columnwidth]
    {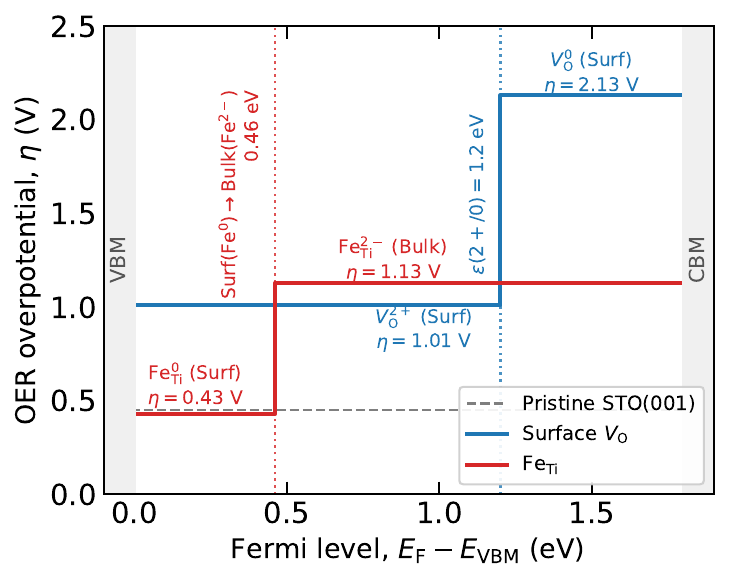}
    \caption{
    Thermodynamic OER overpotential ($\eta$) as a function of the Fermi level
    ($E_{\mathrm{F}} - E_{\mathrm{VBM}}$) inside the \ce{SrTiO3} band gap. Shaded regions indicate the valence band (VBM) and conduction band (CBM).
    The steps show transitions between the most stable defect configurations. At
    $0.46$ eV, Fe dopants undergo a transition where surface
    $\mathrm{Fe}_{\mathrm{Ti}}^{\times}$ ($\eta = 0.43$ V) desegregates into bulk-like
    layers as $\mathrm{Fe}_{\mathrm{Ti}}^{''}$ ($\eta = 1.13$ V). For oxygen vacancies,
    the surface segregated state undergoes a charge transition at $1.2$ eV, switching
    from $V_{\mathrm{O}}^{\bullet\bullet}$ ($\eta = 1.01$ V) to $V_{\mathrm{O}}^{0}$ ($\eta = 2.13$ V).
    }
    \label{fig:fermi_level_oer_crossover}
\end{figure}

Combining the OER results with the defect thermodynamics of
Sec.~\ref{sec:defect_formation} yields a Fermi-level-dependent picture of
the catalytic activity, summarized in
Fig.~\ref{fig:fermi_level_oer_crossover}. For each value of
$E_\mathrm{F}$, the thermodynamically stable defect configuration (charge
state and location) is selected from
Fig.~\ref{fig:defect_formation_ctls}, and its overpotential is plotted.

For Fe, the stable configuration near the VBM
($E_\mathrm{F} - E_\mathrm{VBM} < 0.46$~eV) is the surface-segregated
\FeTix{} with $\eta = 0.43$~V, slightly below the pristine value. Above
$0.46$~eV, a combined charge-state and segregation transition occurs: Fe
desegregates into bulk-like layers as \FeTipp{}, and the overpotential of
the stable configuration rises to $1.13$~V. For oxygen vacancies, the
surface-segregated state is stable across the entire gap, but its charge
state switches at $E_\mathrm{F} - E_\mathrm{VBM} = 1.20$~eV from \VOpp{}
($\eta = 1.01$~V) to \VOx{} ($\eta = 2.13$~V), the latter overstabilizing
*OH and *O. Even though surface segregation of \VO{} is favored
thermodynamically, it is detrimental catalytically relative to the
bulk-like \VOpp{} case ($\eta = 0.45$~V).

We emphasize that this idealized picture must be transferred to real materials with care. The Fermi level is not a free parameter. It is fixed self-consistently by global charge neutrality, set by the coupled ionization of all donor- and acceptor-like species. In the present case, the donor \VO{} and the acceptor \FeTi{} tend to compensate and pin $E_\mathrm{F}$. The Fermi level is further influenced by defect association, applied bias, pH, and oxygen partial pressure under operating conditions. Likewise, the realized defect distribution is governed not by thermodynamics alone. Cation and vacancy diffusion barriers, together with processing and cooling history, determine whether the equilibrium segregation profile is reached or a metastable, kinetically frozen distribution is retained. Nevertheless, Fig.~\ref{fig:fermi_level_oer_crossover} delineates which defect charge states and locations are catalytically benign and which are detrimental, and how this assignment shifts with $E_\mathrm{F}$.

\section{Summary and Conclusion}
In this work we have studied the OER on STO as a prototypical
reaction/material system to investigate the tunability of surface
reactivity by means of Fermi-level engineering. To this end, we have
calculated defect formation energies and charge transition levels for
oxygen vacancies and Fe dopants in multiple charge states in bulk-like
and surface sites, and then evaluated OER free-energy profiles in the
presence of these defects. Oxygen vacancies segregate to the surface
across the entire band gap ($\Delta E_\mathrm{seg} = -0.50$ to
$-0.80$~eV), whereas for \FeTi{} surface segregation is favorable only
for \FeTix{} and only for $E_\mathrm{F} - E_\mathrm{VBM} \lesssim
0.46$~eV. The chosen defects generally increase the OER overpotential
relative to the pristine surface ($\eta = 0.45$~V); the only exceptions
are bulk-like \VOpp{} and \FeTix{} (regardless of location), which leave
the overpotential essentially unchanged at $0.43$--$0.48$~V. Although none of the defects studied here improves the OER overpotential relative to the pristine surface, the central result is conceptual. The adsorption energies of the OER intermediates depend on the spatial location and charge state of the defects, and the charge state is in turn set by the Fermi level. The OER overpotential itself therefore becomes a Fermi-level-dependent quantity. \\
Although not directly tunable, the Fermi level is coupled to the defects
present and can be shifted by co-doping, external bias, or the gas-phase
environment~\cite{klein2023fermi}, thus offering a route to stabilize
benign configurations like \FeTix{} and \VOpp{} while avoiding
detrimental ones. This illustrates
the general concept of Fermi-level engineering of OER
overpotentials, which further dopant and
defect-complex screening could turn into practical gains for
\ce{SrTiO3}-based catalysts.

\begin{acknowledgments}
This work was carried out within the Collaborative Research Center FLAIR (Fermi Level Engineering Applied to Oxide Electroceramics), funded by the German Research Foundation (DFG, Project ID 463184206 – SFB 1548). The authors gratefully acknowledge the computing resources provided by the high-performance computing systems of the Paderborn Center for Parallel Computing (PC$^2$) and the Lichtenberg II high-performance computer of TU Darmstadt. These centers are jointly supported by the Federal Ministry of Education and Research (BMBF) and the state governments participating in the NHR4CES.
The authors would like to thank Professor Karsten Albe for his guidance, valuable discussions, and for providing the resources of the Materials Modelling Division.
\end{acknowledgments}
\bibliographystyle{apsrev4-2}
\bibliography{references}

@article{raveendran2023comprehensive,
  title={A comprehensive review on the electrochemical parameters and recent material development of electrochemical water splitting electrocatalysts},
  author={Raveendran, Asha and Chandran, Mijun and Dhanusuraman, Ragupathy},
  journal={RSC advances},
  volume={13},
  number={6},
  pages={3843--3876},
  year={2023},
  publisher={Royal Society of Chemistry}
}

@article{wen2021strain,
  title={Strain effect on oxygen evolution reaction of the SrTiO3 (0 0 1) surface},
  author={Wen, Linyuan and Li, Mingtao and Shi, Jinwen and Liu, Yingzhe and Yu, Tao and Liu, Maochang and Zhou, Zhaohui},
  journal={Applied Physics Letters},
  volume={119},
  number={10},
  year={2021},
  publisher={AIP Publishing}
}

@article{calle2013number,
  title={Number of outer electrons as descriptor for adsorption processes on transition metals and their oxides},
  author={Calle-Vallejo, Federico and Inoglu, Nilay G and Su, Hai-Yan and Martinez, Jose I and Man, Isabela C and Koper, Marc TM and Kitchin, John R and Rossmeisl, Jan},
  journal={Chemical Science},
  volume={4},
  number={3},
  pages={1245--1249},
  year={2013},
  publisher={Royal Society of Chemistry}
}

@article{wrighton1976strontium,
  title={Strontium titanate photoelectrodes. Efficient photoassisted electrolysis of water at zero applied potential},
  author={Wrighton, Mark S and Ellis, Arthur B and Wolczanski, Peter T and Morse, David L and Abrahamson, Harmon B and Ginley, David S},
  journal={Journal of the American Chemical Society},
  volume={98},
  number={10},
  pages={2774--2779},
  year={1976},
  publisher={ACS Publications}
}

@misc{avcioǧluphotocatalytic,
  title={Photocatalytic Overall Water Splitting by SrTiO3: Progress Report and Design Strategies, ACS Appl. Energy Mater. 6 (2023) 1134--1154},
  author={Avc{\i}oǧlu, C and Avc{\i}oǧlu, S and Bekheet, MF and Gurlo, A}
}

@article{zhang2024crystal,
  title={Crystal facet engineering on SrTiO3 enhances photocatalytic overall water splitting},
  author={Zhang, Yang and Wu, Xuefeng and Wang, Zhi-Hao and Peng, Yu and Liu, Yuanwei and Yang, Shuang and Sun, Chenghua and Xu, Xiaoxiang and Zhang, Xie and Kang, Jun and others},
  journal={Journal of the American Chemical Society},
  volume={146},
  number={10},
  pages={6618--6627},
  year={2024},
  publisher={ACS Publications}
}

@article{domen1980photocatalytic,
  title={Photocatalytic decomposition of water vapour on an NiO--SrTiO 3 catalyst},
  author={Domen, Kazunari and Naito, Shuichi and Soma, Mitsuyuki and Onishi, Takaharu and Tamaru, Kenzi},
  journal={Journal of the Chemical Society, Chemical Communications},
  number={12},
  pages={543--544},
  year={1980},
  publisher={Royal Society of Chemistry}
}

@article{grimaud2017activating,
  title={Activating lattice oxygen redox reactions in metal oxides to catalyse oxygen evolution},
  author={Grimaud, Alexis and Diaz-Morales, Oscar and Han, Binghong and Hong, Wesley T and Lee, Yueh-Lin and Giordano, Livia and Stoerzinger, Kelsey A and Koper, Marc TM and Shao-Horn, Yang},
  journal={Nature chemistry},
  volume={9},
  number={5},
  pages={457--465},
  year={2017},
  publisher={Nature Publishing Group UK London}
}

@article{man2011universality,
  title={Universality in oxygen evolution electrocatalysis on oxide surfaces},
  author={Man, Isabela C and Su, Hai-Yan and Calle-Vallejo, Federico and Hansen, Heine A and Mart{\'\i}nez, Jos{\'e} I and Inoglu, Nilay G and Kitchin, John and Jaramillo, Thomas F and N{\o}rskov, Jens K and Rossmeisl, Jan},
  journal={ChemCatChem},
  volume={3},
  number={7},
  pages={1159--1165},
  year={2011},
  publisher={Wiley Online Library}
}

@article{xie2022oxygen,
  title={Oxygen evolution reaction in alkaline environment: material challenges and solutions},
  author={Xie, Xiaohong and Du, Lei and Yan, Litao and Park, Sehkyu and Qiu, Yang and Sokolowski, Joshua and Wang, Wei and Shao, Yuyan},
  journal={Advanced Functional Materials},
  volume={32},
  number={21},
  pages={2110036},
  year={2022},
  publisher={Wiley Online Library}
}

@article{li2019recent,
  title={Recent advances in the development of water oxidation electrocatalysts at mild pH},
  author={Li, Peipei and Zhao, Runbo and Chen, Hongyu and Wang, Huanbo and Wei, Peipei and Huang, Hong and Liu, Qian and Li, Tingshuai and Shi, Xifeng and Zhang, Youyu and others},
  journal={Small},
  volume={15},
  number={13},
  pages={1805103},
  year={2019},
  publisher={Wiley Online Library}
}

@article{kresse1996efficient,
  title={Efficient iterative schemes for ab initio total-energy calculations using a plane-wave basis set},
  author={Kresse, Georg and Furthm{\"u}ller, J{\"u}rgen},
  journal={Physical review B},
  volume={54},
  number={16},
  pages={11169},
  year={1996},
  publisher={APS}
}

@article{kresse1996efficiency,
  title={Efficiency of ab-initio total energy calculations for metals and semiconductors using a plane-wave basis set},
  author={Kresse, Georg and Furthm{\"u}ller, J{\"u}rgen},
  journal={Computational materials science},
  volume={6},
  number={1},
  pages={15--50},
  year={1996},
  publisher={Elsevier}
}

@article{blochl1994projector,
  title={Projector augmented-wave method},
  author={Bl{\"o}chl, Peter E},
  journal={Physical review B},
  volume={50},
  number={24},
  pages={17953},
  year={1994},
  publisher={APS}
}

@article{kresse1999ultrasoft,
  title={From ultrasoft pseudopotentials to the projector augmented-wave method},
  author={Kresse, Georg and Joubert, Daniel},
  journal={Physical review b},
  volume={59},
  number={3},
  pages={1758},
  year={1999},
  publisher={APS}
}

@article{monkhorst1976special,
  title={Special points for Brillouin-zone integrations},
  author={Monkhorst, Hendrik J and Pack, James D},
  journal={Physical review B},
  volume={13},
  number={12},
  pages={5188},
  year={1976},
  publisher={APS}
}

@article{cui2019oxygen,
  title={Oxygen evolution reaction (OER) on clean and oxygen deficient low-index SrTiO3 surfaces: a theoretical systematic study},
  author={Cui, Mengsi and Liu, Taifeng and Li, Qiuye and Yang, Jianjun and Jia, Yu},
  journal={ACS Sustainable Chemistry \& Engineering},
  volume={7},
  number={18},
  pages={15346--15353},
  year={2019},
  publisher={ACS Publications}
}

@article{sokolov2024computational,
  title={Computational study of oxygen evolution reaction on flat and stepped surfaces of strontium titanate},
  author={Sokolov, Maksim and Mastrikov, Yuri A and Bocharov, Dmitry and Krasnenko, Veera and Zvejnieks, Guntars and Exner, Kai S and Kotomin, Eugene A},
  journal={Catalysis Today},
  volume={432},
  pages={114609},
  year={2024},
  publisher={Elsevier}
}

@article{wang2006oxidation,
  title   = {Oxidation energies of transition metal oxides within the {GGA}+{U} framework},
  author  = {Wang, Lei and Maxisch, Thomas and Ceder, Gerbrand},
  journal = {Phys. Rev. B},
  volume  = {73},
  pages   = {195107},
  year    = {2006},
  doi     = {10.1103/PhysRevB.73.195107}
}

@article{wang2021framework,
  title   = {A framework for quantifying uncertainty in {DFT} energy corrections},
  author  = {Wang, Amanda and Kingsbury, Ryan and McDermott, Matthew and Horton, Matthew and Jain, Anubhav and Ong, Shyue Ping and Dwaraknath, Shyam and Persson, Kristin A.},
  journal = {Sci. Rep.},
  volume  = {11},
  pages   = {15496},
  year    = {2021},
  doi     = {10.1038/s41598-021-94550-5}
}

@article{rothschild2006electronic,
  author  = {Rothschild, Avner and Menesklou, Wolfgang and Tuller, Harry L. and Ivers-Tiff{\'e}e, Ellen},
  title   = {Electronic Structure, Defect Chemistry, and Transport Properties of {SrTi$_{1-x}$Fe$_x$O$_{3-y}$} Solid Solutions},
  journal = {Chem. Mater.},
  volume  = {18},
  pages   = {3651--3659},
  year    = {2006},
  doi     = {10.1021/cm052803x}
}

@article{hayden2018oxygen,
  author  = {Hayden, Brian E. and Rogers, Fiona K.},
  title   = {Oxygen Reduction and Oxygen Evolution on {SrTi$_{1-x}$Fe$_x$O$_{3-y}$} ({STFO}) Perovskite Electrocatalysts},
  journal = {J. Electroanal. Chem.},
  volume  = {819},
  pages   = {275--282},
  year    = {2018},
  doi     = {10.1016/j.jelechem.2017.10.056}
}

@article{zhang2019high,
  author  = {Zhang, Shan-Lin and Cox, Dalton and Yang, Hao and Park, Beom-Kyeong and Li, Cheng-Xin and Li, Chang-Jiu and Barnett, Scott A.},
  title   = {High Stability {SrTi$_{1-x}$Fe$_x$O$_{3-\delta}$} Electrodes for Oxygen Reduction and Oxygen Evolution Reactions},
  journal = {J. Mater. Chem. A},
  volume  = {7},
  pages   = {21447--21458},
  year    = {2019},
  doi     = {10.1039/C9TA07548H}
}

@article{merkle2008how,
  author  = {Merkle, Rotraut and Maier, Joachim},
  title   = {How Is Oxygen Incorporated into Oxides? {A} Comprehensive Kinetic Study of a Simple Solid-State Reaction with {SrTiO$_3$} as a Model Material},
  journal = {Angew. Chem. Int. Ed.},
  volume  = {47},
  pages   = {3874--3894},
  year    = {2008},
  doi     = {10.1002/anie.200700987}
}

@article{freysoldt2014first,
  author  = {Freysoldt, Christoph and Grabowski, Blazej and Hickel, Tilmann and Neugebauer, J{\"o}rg and Kresse, Georg and Janotti, Anderson and Van de Walle, Chris G.},
  title   = {First-Principles Calculations for Point Defects in Solids},
  journal = {Rev. Mod. Phys.},
  volume  = {86},
  pages   = {253--305},
  year    = {2014},
  doi     = {10.1103/RevModPhys.86.253}
}

@article{klein2023fermi,
  title={The Fermi energy as common parameter to describe charge compensation mechanisms: A path to Fermi level engineering of oxide electroceramics},
  author={Klein, Andreas and Albe, Karsten and Bein, Nicole and Clemens, Oliver and Creutz, Kim Alexander and Erhart, Paul and Frericks, Markus and Ghorbani, Elaheh and Hofmann, Jan Philipp and Huang, Binxiang and others},
  journal={Journal of Electroceramics},
  volume={51},
  number={3},
  pages={147--177},
  year={2023},
  publisher={Springer}
}

@article{lankauf2021effect,
  title={The effect of Fe on chemical stability and oxygen evolution performance of high surface area SrTix-1FexO3-$\delta$ mixed ionic-electronic conductors in alkaline media},
  author={Lankauf, Krystian and Mrozi{\'n}ski, Aleksander and B{\l}aszczak, Patryk and G{\'o}rnicka, Karolina and Ignaczak, Justyna and {\L}api{\'n}ski, Marcin and Karczewski, Jakub and Cempura, Grzegorz and Jasi{\'n}ski, Piotr and Molin, Sebastian},
  journal={International Journal of Hydrogen Energy},
  volume={46},
  number={56},
  pages={28575--28590},
  year={2021},
  publisher={Elsevier}
}

@article{kubacki2018impact,
  title={Impact of Fe doping on the electronic structure of SrTiO3 thin films determined by resonant photoemission},
  author={Kubacki, J and Kajewski, D and Goraus, J and Szot, K and Koehl, A and Lenser, Ch and Dittmann, R and Szade, J},
  journal={The journal of chemical physics},
  volume={148},
  number={15},
  year={2018},
  publisher={AIP Publishing}
}

@article{zhou2011effect,
  title={Effect of metal doping on electronic structure and visible light absorption of SrTiO3 and NaTaO3 (Metal= Mn, Fe, and Co)},
  author={Zhou, Xin and Shi, Jingying and Li, Can},
  journal={The Journal of Physical Chemistry C},
  volume={115},
  number={16},
  pages={8305--8311},
  year={2011},
  publisher={ACS Publications}
}

@article{qin2021codoped,
  title={La, Al-codoped SrTiO3 as a photocatalyst in overall water splitting: significant surface engineering effects on defect engineering},
  author={Qin, Yalei and Fang, Fan and Xie, Zhengzheng and Lin, Huiwen and Zhang, Kai and Yu, Xu and Chang, Kun},
  journal={ACS Catalysis},
  volume={11},
  number={18},
  pages={11429--11439},
  year={2021},
  publisher={ACS Publications}
}

@article{zhao2019electronic,
  title={Electronic structure basis for enhanced overall water splitting photocatalysis with aluminum doped SrTiO 3 in natural sunlight},
  author={Zhao, Zeqiong and Goncalves, Renato V and Barman, Sajib K and Willard, Emma J and Byle, Edaan and Perry, Russell and Wu, Zongkai and Huda, Muhammad N and Moul{\'e}, Adam J and Osterloh, Frank E},
  journal={Energy \& Environmental Science},
  volume={12},
  number={4},
  pages={1385--1395},
  year={2019},
  publisher={Royal Society of Chemistry}
}

@article{w683-tvc4,
  title = {Instability of $AB{\mathrm{O}}_{3}$ perovskite surfaces induced by vacancy formation ($A=\mathrm{Ca},\mathrm{Sr},\mathrm{Ba}$; $B=\mathrm{Ti},\mathrm{Zr},\mathrm{Sn}$)},
  author = {Taylor, Ned Thaddeus and Morgan, Michael Thomas and Hepplestone, Steven Paul},
  journal = {Phys. Rev. B},
  volume = {112},
  issue = {4},
  pages = {045308},
  numpages = {14},
  year = {2025},
  month = {Jul},
  publisher = {American Physical Society},
  doi = {10.1103/w683-tvc4},
  url = {https://link.aps.org/doi/10.1103/w683-tvc4}
}

@misc{perdew1996generalized,
  title={Generalized gradient approximation made simple, PhysRevLett 77 (1996) 3865--3868},
  author={Perdew, JP and Burke, K and Ernzerhof, M},
  year={1996}
}

@article{freysoldt2009fully,
  title={Fully ab initio finite-size corrections for charged-defect supercell calculations},
  author={Freysoldt, Christoph and Neugebauer, J{\"o}rg and Van de Walle, Chris G},
  journal={Physical review letters},
  volume={102},
  number={1},
  pages={016402},
  year={2009},
  publisher={APS}
}

@article{freysoldt2018first,
  title={First-principles calculations for charged defects at surfaces, interfaces, and two-dimensional materials in the presence of electric fields},
  author={Freysoldt, Christoph and Neugebauer, J{\"o}rg},
  journal={Physical Review B},
  volume={97},
  number={20},
  pages={205425},
  year={2018},
  publisher={APS}
}

@article{Lewis2006,
  author  = {Lewis, Nathan S. and Nocera, Daniel G.},
  title   = {Powering the Planet: Chemical Challenges in Solar Energy Utilization},
  journal = {Proc. Natl. Acad. Sci. U.S.A.},
  volume  = {103},
  pages   = {15729--15735},
  year    = {2006},
  doi     = {10.1073/pnas.0603395103}
}

@article{liu2024perovskite,
  title={Perovskite oxides toward oxygen evolution reaction: intellectual design strategies, properties and perspectives},
  author={Liu, Lin-Bo and Yi, Chenxing and Mi, Hong-Cheng and Zhang, Song Lin and Fu, Xian-Zhu and Luo, Jing-Li and Liu, Subiao},
  journal={Electrochemical Energy Reviews},
  volume={7},
  number={1},
  pages={14},
  year={2024},
  publisher={Springer}
}

@article{Seh2017,
  author  = {Seh, Zhi Wei and Kibsgaard, Jakob and Dickens, Colin F. and Chorkendorff, Ib and N{\o}rskov, Jens K. and Jaramillo, Thomas F.},
  title   = {Combining Theory and Experiment in Electrocatalysis: Insights into Materials Design},
  journal = {Science},
  volume  = {355},
  pages   = {eaad4998},
  year    = {2017},
  doi     = {10.1126/science.aad4998}
}

@article{Rossmeisl2007,
  author  = {Rossmeisl, Jan and Qu, Zhi-Wei and Zhu, Hao and Kroes, Geert-Jan and N{\o}rskov, Jens K.},
  title   = {Electrolysis of Water on Oxide Surfaces},
  journal = {J. Electroanal. Chem.},
  volume  = {607},
  pages   = {83--89},
  year    = {2007},
  doi     = {10.1016/j.jelechem.2006.11.008}
}

@article{Suen2017,
  author  = {Suen, Nian-Tzu and Hung, Sung-Fu and Quan, Quan and Zhang, Nan and Xu, Yi-Jun and Chen, Hao Ming},
  title   = {Electrocatalysis for the Oxygen Evolution Reaction: Recent Development and Future Perspectives},
  journal = {Chem. Soc. Rev.},
  volume  = {46},
  pages   = {337--365},
  year    = {2017},
  doi     = {10.1039/C6CS00328A}
}

@article{Man2011,
  author  = {Man, Isabela C. and Su, Hai-Yan and Calle-Vallejo, Federico and Hansen, Heine A. and Mart{\'\i}nez, Jos{\'e} I. and Inoglu, Nilay G. and Kitchin, John and Jaramillo, Thomas F. and N{\o}rskov, Jens K. and Rossmeisl, Jan},
  title   = {Universality in Oxygen Evolution Electrocatalysis on Oxide Surfaces},
  journal = {ChemCatChem},
  volume  = {3},
  pages   = {1159--1165},
  year    = {2011},
  doi     = {10.1002/cctc.201000397}
}

@article{Hwang2017,
  author  = {Hwang, Jonathan and Rao, Reshma R. and Giordano, Livia and Katayama, Yu and Yu, Yang and Shao-Horn, Yang},
  title   = {Perovskites in Catalysis and Electrocatalysis},
  journal = {Science},
  volume  = {358},
  pages   = {751--756},
  year    = {2017},
  doi     = {10.1126/science.aam7092}
}

@article{Suntivich2011,
  author  = {Suntivich, Jin and May, Kevin J. and Gasteiger, Hubert A. and Goodenough, John B. and Shao-Horn, Yang},
  title   = {A Perovskite Oxide Optimized for Oxygen Electrocatalysis in Alkaline Media},
  journal = {Science},
  volume  = {334},
  pages   = {1383--1385},
  year    = {2011},
  doi     = {10.1126/science.1212858}
}

@article{Akbashev2018,
  author  = {Akbashev, Alexie R. and Zhang, Liang and Mefford, J. Tyler and Park, Jihun and Adler, Stuart B. and Tsai, Haw-Wen and Chueh, William C.},
  title   = {Activation of Ultrathin {SrTiO$_3$} with Subsurface {SrRuO$_3$} for the Oxygen Evolution Reaction},
  journal = {Energy Environ. Sci.},
  volume  = {11},
  pages   = {1762--1769},
  year    = {2018},
  doi     = {10.1039/C8EE00210J}
}

@article{Mefford2016,
  author  = {Mefford, J. Tyler and Rong, Xi and Abakumov, Artem M. and Harber, William G. and Ceder, Gerbrand and Chueh, William C.},
  title   = {Water Electrolysis on {La$_{1-x}$Sr$_x$CoO$_{3-\delta}$} Perovskite Electrocatalysts},
  journal = {Nat. Commun.},
  volume  = {7},
  pages   = {11053},
  year    = {2016},
  doi     = {10.1038/ncomms11053}
}

@article{Grimaud2017,
  author  = {Grimaud, Alexis and Diaz-Morales, Oscar and Han, Binghong and Hong, Wesley Tao and Lee, Yueh-Lin and Giordano, Livia and Stoerzinger, Kelsey A. and Koper, Marc T. M. and Shao-Horn, Yang},
  title   = {Activating Lattice Oxygen Redox Reactions in Metal Oxides to Catalyse Oxygen Evolution},
  journal = {Nat. Chem.},
  volume  = {9},
  pages   = {457--465},
  year    = {2017},
  doi     = {10.1038/nchem.2695}
}

@article{Yoo2018,
  author  = {Yoo, Jong Suk and Rong, Xi and Liu, Yusu and Kolpak, Alexie M.},
  title   = {Role of Lattice Oxygen Participation in Understanding Trends in the Oxygen Evolution Reaction on Perovskites},
  journal = {ACS Catal.},
  volume  = {8},
  pages   = {4628--4636},
  year    = {2018},
  doi     = {10.1021/acscatal.8b00612}
}

@article{Norskov2004,
  author  = {N{\o}rskov, Jens K. and Rossmeisl, Jan and Logadottir, Ashildur and Lindqvist, Lars and Kitchin, John R. and Bligaard, Thomas and J{\'o}nsson, Hannes},
  title   = {Origin of the Overpotential for Oxygen Reduction at a Fuel-Cell Cathode},
  journal = {J. Phys. Chem. B},
  volume  = {108},
  pages   = {17886--17892},
  year    = {2004},
  doi     = {10.1021/jp047349j}
}

\end{document}